\documentclass[]{ceurart}

\usepackage{listings}
\usepackage{xcolor}
\usepackage{microtype}
\newcommand{\code}[1]{\colorbox{gray!15}{\textcolor{black}{\texttt{\textls[100]{#1}}}}}

\usepackage{url}

\usepackage{mdframed}

\newenvironment{llmresponse}{%
    \begin{mdframed}[backgroundcolor=blue!5, linewidth=0pt, innerleftmargin=15pt, innerrightmargin=15pt, innertopmargin=10pt, innerbottommargin=10pt]
    \small\sffamily\setlength{\parindent}{0pt}
}{%
    \end{mdframed}
}

\usepackage{listings}

\lstnewenvironment{codeblock}[1][]{
    \lstset{
        basicstyle=\small\ttfamily,
        backgroundcolor=\color{gray!10},
        frame=leftline,
        framerule=3pt,
        rulecolor=\color{gray!50},
        xleftmargin=15pt,
        xrightmargin=10pt,
        aboveskip=10pt,
        belowskip=5pt, % reduced bottom skip
        breaklines=true
    }
}{
    \ifx\relax#1\relax\else
        \nopagebreak
        \vspace{0.05em}
        \nopagebreak
        \begin{center}
            \small\textit{#1}
        \end{center}
        \vspace{0.3em}
    \fi 
}

\usepackage{enumitem}

\newenvironment{experiment}[1]{%
    \vspace{1em}
    \noindent\textbf{#1}
    \vspace{0.1em}
    \begin{list}{}{%
        \setlength{\leftmargin}{1.5cm}%
        \setlength{\rightmargin}{0pt}%
        \setlength{\parsep}{0.1em}%
        \setlength{\itemsep}{0.1em}%
        \setlength{\topsep}{0pt}%
        \setlength{\partopsep}{0pt}%
        \setlength{\labelwidth}{0pt}%
        \setlength{\labelsep}{0pt}%
    }
    \item[]
}{%
    \end{list}
    \vspace{1em}
}

\newcommand{\expitem}[2]{\item[] \textbf{#1:} #2\vspace{0.1em}}

\begin{document}

\copyrightyear{2025}
\copyrightclause{Copyright for this paper by its authors.
  Use permitted under Creative Commons License Attribution 4.0
  International (CC BY 4.0).}

\conference{\textbf{Preprint}, submitted to the 2nd Workshop on Computational Drama Analysis at DraCor Summit 2025, September 03, 2025, Berlin, Germany}

\title{Agentic DraCor and the Art of Docstring Engineering}
\subtitle{Evaluating MCP-empowered LLM Usage of the DraCor API}

\author[1]{Peer Trilcke}[%
orcid=0000-0002-1421-4320,
email=trilcke@uni-potsdam.de
]
\cormark[1]
\address[1]{University of Potsdam, Germany}
\address[2]{Freie Universität Berlin, Germany}

\author[1]{Ingo Börner}[%
orcid= 0000-0001-8294-2541,
email=ingo.boerner@uni-potsdam.de]

\author[1]{Henny Sluyter-Gäthje}[%
orcid=0000-0003-2969-3237,
email=sluytergaeth@uni-potsdam.de]

\author[1]{Daniil Skorinkin}[%
orcid=0000-0002-1845-9974,
email=daniil.skorinkin@uni-potsdam.de]

\author[2]{Frank Fischer}[%
orcid=0000-0003-2419-6629,
email=fr.fischer@fu-berlin.de]

\author[1]{Carsten Milling}[%
orcid=0000-0003-0553-7512,
email=milling@uni-potsdam.de]

\cortext[1]{Corresponding author.}
%\fntext[1]{These authors contributed equally.}

\begin{abstract}
  This paper reports on the implementation and evaluation of a Model Context Protocol (MCP) server for DraCor, enabling Large Language Models (LLM) to autonomously interact with the DraCor API. We conducted experiments focusing on tool selection and application by the LLM, employing a qualitative approach that includes systematic observation of prompts to understand how LLMs behave when using MCP tools, evaluating “Tool Correctness”, “Tool-Calling Efficiency”, and “Tool-Use Reliability”. Our findings highlight the importance of “Docstring Engineering”, defined as reflexively crafting tool documentation to optimize LLM-tool interaction. Our experiments demonstrate both the promise of agentic AI for research in Computational Literary Studies and the essential infrastructure development needs for reliable Digital Humanities infrastructures.
\end{abstract}

\begin{keywords}
  Model Context Protocol \sep
  MCP \sep
  Large Language Model \sep
  LLM \sep
  Claude \sep
  Evaluation \sep
  Computational Literary Studies \sep
  Computational Drama Analysis \sep
  DraCor
\end{keywords}

\maketitle

\begin{flushright}
\textit{there is software that … treats other web sites as just\\ … another function it can call to get things done}
\\Aaron Swartz (2013): A Programmable Web
\end{flushright}

\section{How to Do Things With MCP Servers. Introduction}

\subsection{Making DraCor machine-actionable}

According to its own claim, the API-driven DraCor platform aims to make literary corpora “machine-actionable” \citep[p.~7]{boerner_2023_d71}, showcasing the concept of “Programmable Corpora” (\citealt{fischer_2019_programmable}; \citealt{boerner_2023_d71}; \citealt{boerner_2025_d74}; cf. \citealt{swartz_2013_programmable-web}). That is, machines would interact with corpora through program code, as exemplified by the Python and R wrappers “pydracor” and “rdracor” \citep{sluyter-gaethje_2023_d72}. 

Now, DraCor is gaining a new dimension with the current third wave of AI evolution, known as the transition to “Agentic AI” \citep{huang_2025_agentic}. In the agential paradigm, machines are currently being enabled to more or less autonomously perform concrete actions. A much-promising development in this direction is the “Model Context Protocol” (MCP) introduced by \citet{anthropic_2024_mcp}, which enables a Large Language Model (LLM) to do various things with the help of a server and the functionalities implemented there \citep{hou_2025_mcp}. What can be understood as agential empowerment of LLMs means, in Anthropic's current concept, that MCP enables the LLM to “perform actions through your server.” \citep{anthropic_2025_mcp-intro} In this sense, servers can “expose executable functionality to clients” via the MCP; these functionalities are called “tools.” \citep{anthropic_2025_tools} These “tools are designed to be model-controlled, meaning that tools are exposed from servers to clients with the intention of the AI model being able to automatically invoke them (with a human in the loop to grant approval)” \citep{anthropic_2025_tools}. 

Against the backdrop of this concept of “dynamic tool invocation”, it is the infrastructural expansion in the form of an MCP server that promises to make DraCor “machine-actionable” in the full sense of the word. This will enable LLMs to interact with DraCor and its API more or less autonomously. The way to achieve this is to empower LLMs to use tools specifically tailored to DraCor.

\subsection{An Example}

Let's play through an example: Have you ever asked yourself: “What is the most quantitatively complex German-language play by a female author?”\footnote{Cf. the documentation at \\ \url{https://github.com/dracor-org/dracor-mcp-evaluation/blob/da9d4230bdb36584469bb26eec40469e2694a27e/2025_cda-preprint_documentation/0_DraCor-MCP-Paper_2025-07.md}\label{ftn1}} Probably not. It may even be a research question that no one has ever asked. So let’s pose the question to one of the current generative AI systems, such as ChatGPT or Claude: Usually, the system will perform a series of web searches and give an “interesting” but empirically more or less unfounded answer. The result becomes a matter of interpretation: for example, in a test in mid-July 2025, ChatGPT (“4o”) suggested that Elfriede Jelinek's “Die Schutzbefohlenen” (2013) was the most quantitatively complex play; Claude (“Sonnet 4”), on the other hand, suggested “Sapientia” by Hrotsvitha of Gandersheim (10th century). 

Both models gave good, albeit predominantly qualitative reasons. However, if the question is posed to an agentic empowered LLM (in this case “Claude Sonnet 4”) that can access the DraCor API with the help of tools, something fascinating happens: Once the model – using a tool that allows to request information from the DraCor API \code{corpora} endpoint – has gained an overview of the corpora in DraCor, it searches (analyzing data provided by a different tool for a different API endpoint) for female authors in the German-language drama corpus (“GerDraCor”), then it identifies network metrics that can be understood as indicators of complexity. On this basis, it creates a ranking of 20 plays, sorted according to what the model postulates as a “complexity score” (using the formula \code{network size x ln(edges)}). Using data from the DraCor API, this formula and some further API queries, Claude\footnote{Claude Sonnet 4 (claude-sonnet-4-20250514), Desktop Version (“Claude for Mac”, Version 0.11.6), 2025-07-14, Style: Normal, Extended Thinking: off, Web Search: on, using DraCor MCP Server  \citep{boerner_2025_dracor-mcp}, v1.0.0-alpha.2, \url{https://github.com/dracor-org/dracor-mcp/releases/tag/v1.0.0-alpha.2}. Unfortunately, this version of our MCP server no longer works with the current versions of Claude Desktop (“Claude for Mac”, version 0.12.55). Rapid development cycles of AI tools pose a challenge to reproducibility and may require specific combinations of versions. For current compatibility information and installation instructions, consult the DraCor MCP Server repository README: \url{https://github.com/dracor-org/dracor-mcp}.    
} finally arrives at this result\footnote{The text cited here is from the chat. The documentation of the chat (cf. footnote \ref{ftn1}) differs from this conversational text, though the results are the same. – By the way, despite numerous attempts, we were not able to get the LLM to choose the way of processing and to provide the answer reported here again.}: 

\begin{llmresponse}
[…] 

Based on my comprehensive analysis of the DraCor German Drama Corpus, the most quantitatively complex play written by a woman in German-language drama is "Pfeffer-Rösel oder Die Frankfurter Messe im Jahre 1297" by Charlotte Birch-Pfeiffer.
\\[1em]
Here are the key complexity metrics that establish this play as the most complex:
\\[1em]
\textbf{Network Complexity:}
\begin{itemize}
    \item 40 characters - the highest network size among female-authored plays
    \item 323 network edges - an extremely high number of character interactions
    \item Complexity score: 231.11 (calculated as network size × ln(edges))
\end{itemize}
\textbf{Structural Complexity:}
\begin{itemize}
    \item Average degree: 16.15 (characters interact with many others on average)
    \item Network density: 0.414 (41.4\% of all possible character connections are realized)
    \item Multiple acts and elaborate staging with crowd scenes, imperial court scenes, and complex social hierarchies
\end{itemize}
[…]
\end{llmresponse}

Of course, this conclusion can be debated: The complexity measure used by the agentic enabled LLM only represents a very limited aspect of literary complexity. And neither the play by Jelinek nor the one by Hrotsvitha of Gandersheim, which were suggested by the non-agentic LLMs, are currently part of GerDraCor, for example. And even when we stick to our actual corpus, the answer is wrong: Claude missed the most complex play of a female writer in GerDraCor. With a complexity score of 91 × ln(2010) = 692.14, Lilly Braun’s play “Mutter Maria”\footnote{\url{https://dracor.org/id/ger000573}} (1913) would have been the correct answer according to its own metric.

However, none of these objections is decisive at this point. What is more important is that the approach of the LLM, the strategy it chooses to solve the task and also its answer are committed to the empirical paradigm of CLS – even if the LLM operates erroneously in this paradigm, e.g., because it used an inadequate method to identify all female authors in the corpus. The answers we received from the non-agentic enabled LLMs, on the other hand, primarily follow the hermeneutic-interpretative paradigm of traditional literary studies. 
Against this background, in our research we – at least implicitly – work on and reflect on the possibility that the use of AI architectures (such as an MCP server) could be a way of merging the empirical paradigm that CLS have developed over the past decade with the hermeneutic machine LLMs tend to be.

\subsection{About the following article}

As the example already illustrates, we have recently been experimentally implementing an MCP server for DraCor, inspired by an MCP prototype proposed by Stijn Meijers \citep{meijers_2025_dracor-mcp}. As far as we can see, this is the first ever implementation of an MCP server for Computational Literary Studies (CLS). With a very recent release date in November 2024, it is not surprising that current status descriptions of AI in the CLS do not yet know anything about the MCP (\citealt{bode_2025_cls-ai}, \citealt{akazawa_2025_sprachmodellen}). But even a search for the term “Model Context Protocol” in Google Scholar only returned over 400 hits as of July 2025.

\begin{figure}
    \centering
    \includegraphics[width=1\linewidth]{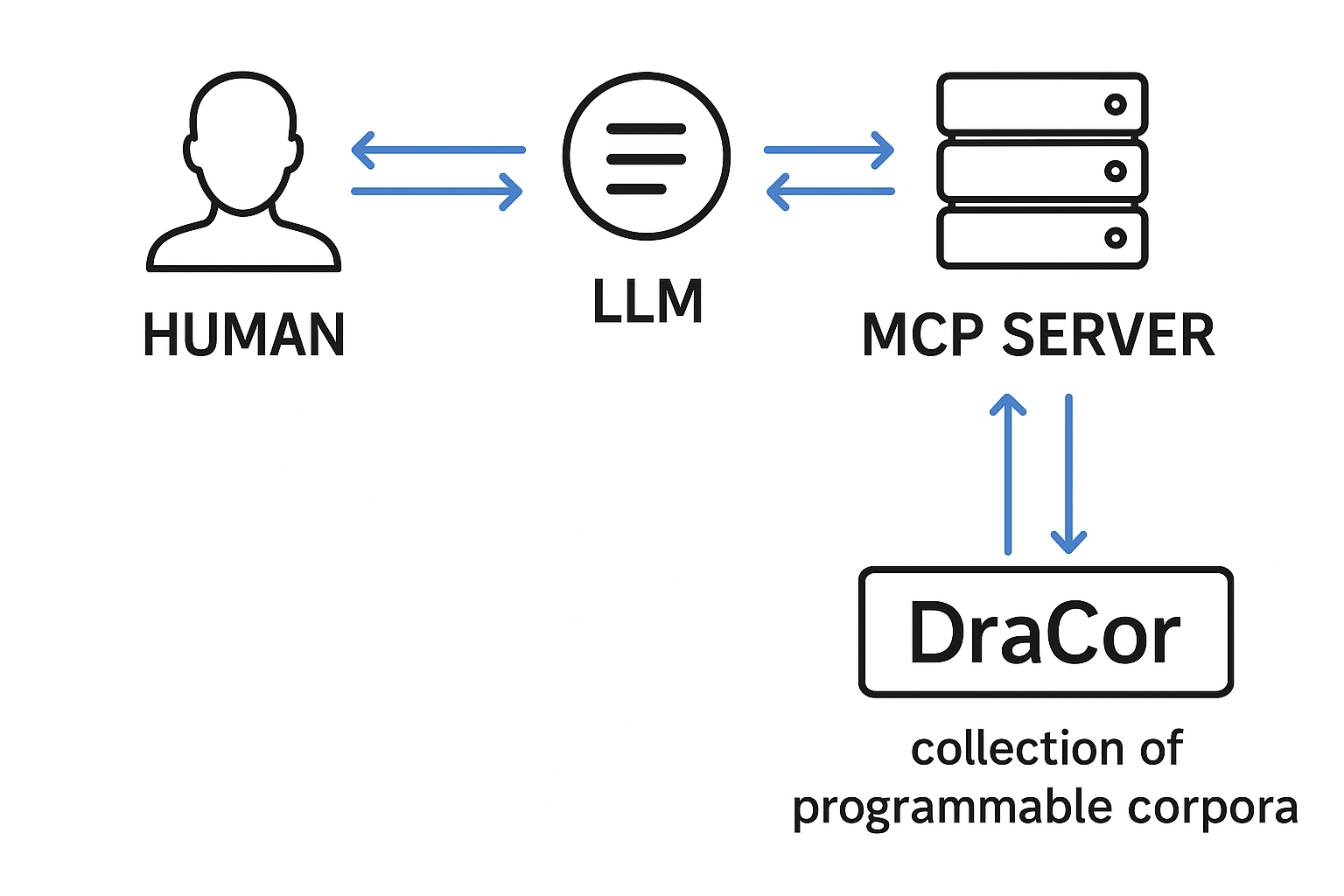}
    \caption{The agentic chain of DraCor MCP pipeline}
    \label{fig:mcp-interaction-schema}
\end{figure}

In the course of this paper, we present first insights from our implementation by describing the DraCor MCP prototype developed by Ingo Börner \citeyear{boerner_2025_dracor-mcp} (chapter \ref{sec:implementation}). We then attempt to evaluate the performance of the interactions between humans, LLM, MCP server and DraCor API (our “agentic chain of actions”, fig. \ref{fig:mcp-interaction-schema}) through a series of experiments (chapter \ref{sec:experiments}). We also want to gain a better understanding of how good these systems are. Where do they go wrong, and what kinds of mistakes do they make?

Our approach to evaluation attempts to take into account the fact that our “agentic chain” is a complex, non-deterministic and, in this respect, potentially fragile and, in any case, multifaceted form of human-machine interaction. In particular, we are interested in how the LLM behaves when using the tools provided by the MCP server: Does it pick the right tool? Does it utilise the tool correctly, efficiently, appropriately? How does it combine tools? As we learned during our experiments, these questions only refer in small parts to aspects of LLM behaviour that can (currently) be formalised and, in the best case, evaluated automatically. On the contrary, it seems more appropriate to pursue an evaluation grounded in something akin to behavioural research: What we have been able to observe and describe so far is, most of all, the interaction of the LLM with our MCP server. This is what we aim to evaluate.

In the light of these considerations, we will only occasionally select a quantitative-formalised approach in the evaluations of the experiments discussed in chapter 3. Instead, we will attempt to develop qualitative-phenomenological descriptions of tool behaviour. One aspect of our phenomenological method is that we will occasionally try to create (through prompts or other parameterisations) slightly varying situations for the tool behaviour of the LLM, mimicking Edmund Husserl's method of “eidetic variation” \citep{Husserl_2012_lehre}.

From the perspective shaped by developing digital tools for computational research using DraCor, particularly the DraCor API, we are led to ask whether we are gaining reliable co-workers in the AI agents that are ever more joining us in our work. Are LLMs, empowered by MCP servers for tool use, perhaps the interface of the future for working with resources such as DraCor? How do LLMs operate with the DraCor API? How do they use tools to work with DraCor? To what extent is the DraCor API becoming a resource predominantly (mis)used by LLMs, rather than one being employed through carefully crafted, purposeful queries?

Before presenting our initial attempts to address these questions from the perspective of research in CLS infrastructures, we have to include two disclaimers. 

First, it is important to emphasize that our test setups depend on various factors: On the capabilities of the LLM, for sure, but at the same time also on the MCP server prototype that we have built (and which could certainly be built differently), on the possibilities and the limitations of the DraCor API – and also on the prompts with which we test, whereby prompt engineering, i.e. the optimization of the prompts, is explicitly not the goal of our test setup.

Second, our technical setup is currently inadequate. While the goal would be to conduct the experiments via API access to an LLM and based on a public MCP server, the following setup, as explained in chapter 2, is based on a locally running (but publicly available) MCP server and the desktop chat interface of Anthropic's Claude model. It is our goal to gradually develop the technological stack further over the next few months and to work on an advanced infrastructure (including, among other things, the possibility of high-frequency re-runs of the experiments and better documentation) for the final version of this paper.

\section{Implementing the DraCor MCP Server}\label{sec:implementation}

The DraCor MCP Server\footnote{The discussed version is 1.0.0-alpha.2, see \url{https://github.com/dracor-org/dracor-mcp/releases/tag/v1.0.0-alpha.2}.} \citep{boerner_2025_dracor-mcp} is implemented using the Python package \code{fastmcp}\footnote{\url{https://gofastmcp.com}}, which provides a framework for creating Model Context Protocol (MCP) servers through decorated Python functions. At its core, the implementation approach emphasizes comprehensive documentation as the primary mechanism for LLM interaction, i.e tool discovery and use. Each tool is essentially a Python function decorated with \code{@mcp.tool()}, where the docstring\footnote{\url{https://peps.python.org/pep-0257/\#what-is-a-docstring}} serves as the crucial interface documentation that provides the LLM with information on the tool's purpose, parameters, and expected usage patterns.

The DraCor MCP Server’s design principle assumes that the quality and comprehensiveness of tool documentation is crucial for the LLM's ability to effectively utilize the available functionality. This approach reflects a broader understanding that in MCP architectures, documentation is not merely auxiliary information but becomes an active part of the system's user interface. The extensive docstrings, combined with type annotations and parameter descriptions, form a semantic contract between the tools and the LLM, enabling autonomous tool selection and parameter passing.

Beyond the 47 core tools currently defined, the DraCor MCP server also defines two resources, though the current implementation primarily focuses on tools due to the limitations of Anthropic’s Claude Desktop\footnote{\url{https://claude.ai/download}} interface. Two key resources are maintained: a list of corpora fetched via Distributed Text Services (DTS, cf. \citealt{Almas_2023_dts}), (\code{DraCor:resource\_corpora\_via\_dts}) and a corpus registry JSON file\footnote{\url{https://github.com/dracor-org/dracor-registry/blob/main/corpora.json}} containing comprehensive metadata about all available corpora, including their development status and availability (\code{DraCor:resource\_corpus\_registry}). In Claude Desktop the resources need to be manually added to the context of the chat.

\subsection{Direct API Endpoint Wrappers}\label{sec:api-endpoint-wrappers}

The largest category consists of tools that provide direct wrappers around DraCor API endpoints, offering unfiltered access to the underlying REST API functionality. These tools maintain a one-to-one correspondence with API endpoints, preserving the full data structure and metadata provided by the DraCor system. Examples include \code{DraCor:get\_corpus}, \code{DraCor:get\_play\_metadata}, \code{DraCor:get\_play\_characters}, and \code{DraCor:get\_spoken\_text}. Similarly, Wikidata integration tools like \code{DraCor:get\_plays\_with\_characters\_by\_wikidata\_id} and \code{DraCor:get\_author\_info\_from\_wikidata} provide access to external identifier-based search and author information retrieval.

A subset of tools in this category provides selective access to specific API endpoints rather than complete endpoint wrapping. The DTS integration tools such as\\ \code{DraCor:get\_citable\_units\_via\_dts} and\\ \code{DraCor:get\_plaintext\_of\_citable\_unit\_via\_dts} exemplify this approach, offering access to fine-grained textual segments, e.g. acts and scenes, and citation capabilities while implementing selective exposure of the DTS endpoint functionality. This category of tools represents the foundational layer of functionality, ensuring that all core DraCor capabilities remain accessible through the MCP interface.

\subsection{Helper Tools}\label{sec:helper-tools}

A significant subset of tools addresses the practical limitations of working with large-scale corpus data in LLM contexts. These “helper tools” implement data filtering, chunking, and pagination strategies. Tools such as \code{DraCor:get\_corpus\_metadata\_paged\_helper},\\ \code{DraCor:get\_minimal\_data\_of\_plays\_of\_corpus\_helper}, and\\ \code{DraCor:get\_corpus\_contents\_paged\_helper} exemplify this category.

The design strategy behind these tools recognizes that simple data retrieval strategies often fail when working with corpora containing hundreds of plays and extensive metadata, e.g. the “metadata table”\footnote{\url{https://dracor.org/api/v1/corpora/fre/metadata}} for corpora like FreDraCor. The helper tools implement flexible selection of batch sizes and prioritize essential metadata fields. In the docstring, annotations include a reference to the respective API wrapper tool and vice versa.

\begin{codeblock}[Listing 1: Docstring of the function that implements the tool \code{DraCor:get\_corpus\_metadata}]
   """Get extended metadata of all plays in a corpus
  
   Data is retrieved from the endpoint /corpora/{corpusname}/metadata
   If the data on the plays does not fit into the context use the tool `get_corpus_metadata_paged_helper` instead,
   which allows for retrieving metadata on the plays in batches.
  
   Args:
       corpus_name (str): Identifier of a corpus, e.g. `ger`, `rus`, `als`
   """ 
\end{codeblock}

\begin{codeblock}[Listing 2: Docstring of the function that implements the tool \code{DraCor:get\_corpus\_metadata\_paged\_helper}]
   """Get metadata on all plays in a corpus in batches


   Data is retrieved from the endpoint /corpora/{corpusname}/metadata, but in batches.
   This normally times out; that is a problem of the API
  
   Args:
       corpus_name (str): Identifier of a corpus, e.g. `ger`, `rus`, `als`
       items_per_page (int): Number of play metadata to retrieve in a batch. Defaults to 50.
       page (int): Number of page of the results to retrieve in a batch request. Defaults to 1 - the first 50 plays.
   """ 
\end{codeblock}

Interestingly, despite the theoretical usefulness of these pagination capabilities, empirical observations suggest that language models rarely make use of batch request features. Instead, they tend to prefer processing complete datasets when possible, and often fail beyond recovery. We will come back to this later.

\subsection{Search and Discovery Tools}\label{sec:search-tools}

The DraCor API's limited native search functionality necessitated the implementation of specialized search tools within the DraCor MCP server. Tools such as\\ \code{DraCor:get\_plays\_in\_corpus\_by\_author\_helper},\\ \code{get\_plays\_in\_corpus\_by\_title\_helper}, and\\ \code{DraCor:get\_plays\_in\_corpus\_by\_year\_normalized} provide essential discovery capabilities that extend beyond the basic API functionality. The implementation assumes that the LLM will frequently need to locate specific plays based on partial information, author names or titles of works rather than the exact DraCor identifiers. This category of tools tries to transform the DraCor system from a purely identifier-based access system into a discovery platform based, among others, on entities.

\subsection{Administrative and Database Management Tools}\label{sec:admin-tools}

The DraCor MCP server includes a complete set of administrative tools that provide write access to the underlying eXist-db database, enabling corpus management and the addition and removal of plays. Tools such as \code{DraCor:add\_corpus}, \code{DraCor:load\_corpus\_from\_repository}, \code{DraCor:add\_play\_to\_corpus}, and \code{DraCor:remove\_play\_from\_corpus} implement the full range of database modification operations available via the DraCor API. Additionally, this category includes quality control tools such as \code{DraCor:validate\_xml\_file}, which serves as a helper for ensuring that content meets DraCor's encoding standards before uploading it to local instances. 

While these administrative capabilities are typically restricted in production environments, they prove invaluable for local development and research environments, particularly when used in conjunction with Docker-based DraCor instances \citep[pp. 25–31]{boerner_2024_d73}. The implementation of these tools within the MCP framework demonstrates the potential for LLM-mediated database administration, though it also raises important considerations about access control and operation safety that must be carefully managed in production deployments.

\subsection{Documentation and System Information Tools}\label{sec:documentation-tools}

A set of tools provides access to various resources of DraCor system documentation, enabling the LLM to access and relay information about the system's capabilities, data structures, and encoding standards. Tools such as \code{DraCor:get\_api\_feature\_list}, \code{get\_openapi\_specification}, \code{get\_table\_of\_contents\_from\_odd} fall into this category.

This documentation-focused approach recognizes that effective use of a complex Digital Humanities system requires not just data access but also contextual understanding of the data structures, encoding practices, and analytical possibilities (cf. \citealt{boerner_2025_explainability}) By making documentation programmatically accessible, these tools enable dynamic help systems, i.e. supporting users in encoding TEI files, understanding API responses, etc.

\subsection{Frontend Access and DraCor Research Integration}\label{sec:frontend-tools}

A distinctive category of tools provides access to information and functionalities that are primarily available through the DraCor frontend interface, bridging the gap between the API-driven MCP server and the web-based user interface. Two tools exemplify this category: \code{DraCor:get\_dracor\_based\_research} and \\ \code{DraCor:get\_links\_to\_playdata\_helper}.

The \code{DraCor:get\_dracor\_based\_research} tool provides programmatic access to the comprehensive list of publications that utilize DraCor resources in their research, information that is typically accessed through the DraCor website's research documentation.\footnote{\url{https://dracor.org/doc/research}. The MCP Server provides access to the underlying YAML file \url{https://github.com/dracor-org/dracor-frontend/blob/main/public/doc/research.md}.} This tool enables the LLM to provide users with relevant academic context and citations, supporting literature reviews and helping researchers understand how DraCor data has been employed in scholarly work.

The \code{get\_links\_to\_playdata\_helper} tool generates direct links to various DraCor frontend features and external analytical tools, including the Download Tab (\citealt[pp. 75–76]{boerner_2023_d71}) for accessing play data in different formats, and the Tool Tab (\citealt[pp. 103–106]{boerner_2025_d74}) which provides connections to external services such as the CLARIN Language Resource Switchboard, Voyant Tools, and Gephi Lite. This tool effectively extends the MCP server's capabilities beyond direct data access to include guided access to the broader DraCor ecosystem and its integrated external tools.

These tools recognize that while the MCP server provides comprehensive API access, the DraCor frontend offers additional visualization, analysis, and export capabilities that complement the programmatic access to the data via the API. By providing structured access to these frontend features, the tools enable seamless workflow integration between conversational AI interaction and traditional web-based digital humanities tools.

\subsection{A (First) Step Back: Some Challenges}

Several significant implementation challenges shaped the final architecture of the DraCor MCP Server. The most prominent challenge involves managing the tension between comprehensive data access and practical usability within limited LLM context windows and processing constraints. The solution involved implementing multiple access strategies for the same underlying data: direct access tools for complete datasets, filtered tools for essential information only, and paginated tools for incremental access.

Another critical design decision involved the balance between tool specificity and generality. While highly specific tools provide optimal user experience for common tasks, they also make tool selection for the LLM much harder. The implemented solution favors a moderate level of tool specificity, providing dedicated tools for common operations while maintaining more general tools for edge cases and exploratory workflows.

The administrative tools present perhaps the most complex implementation challenge, requiring careful balance between functionality and safety. The current implementation assumes a trusted environment with appropriate access controls managed at the infrastructure level, but production deployments would require additional authentication and authorization mechanisms integrated into the MCP framework itself.

\section{Experiments}\label{sec:experiments}

As the documentation in Chapter 2 makes clear, the DraCor MCP Server prototype we developed is a multi-faceted test system that for example also includes admin tools for the DraCor ecosystem. In the following evaluation (which will certainly be the first in a whole series of evaluations), we want to focus on those tools that directly enable the LLM to interact autonomously with DraCor's knowledge resources, i.e. those that are potentially most relevant for research.  In our experiments, we therefore concentrate on prompting queries that we expect will trigger the use of tools from the groups “direct API endpoint wrapper” (chapter \ref{sec:api-endpoint-wrappers}), “helper tools” (chapter \ref{sec:helper-tools}) and “search and discovery tools” (chapter \ref{sec:search-tools}). For all experiments, we used the setup already mentioned above,\footnote{Claude Sonnet 4 (claude-sonnet-4-20250514), Desktop Version (“Claude for Mac”, Version 0.11.6), 2025-07-14, Style: Normal, Extended Thinking: off, Web Search: on, using DraCor MCP Server \citep{boerner_2025_dracor-mcp}, v1.0.0-alpha.2, \url{https://github.com/dracor-org/dracor-mcp/releases/tag/v1.0.0-alpha.2}.} with Claude Sonnet 4 as LLM.

As initial explorations have shown that even queries of medium complexity can lead to very long sequences of tool invocations (which on the one hand are very demanding to evaluate and on the other hand often cannot be evaluated at all because the maximum length for conversations in the LLM is reached), we have – for this very first evaluation – tried to formulate queries that are as simple as possible. At the same time, we wanted the queries to cover a certain spectrum of research questions, epistemic objects to which they are directed and, finally, ways of writing prompts. Against this background, we determined that we wanted to learn about and evaluate (and in a sense: look at the essence of) LLM's ability to use tools through queries, which we (eidetically) varied with regard to the following aspects:

\begin{itemize}
    \item \textbf{scope}: we target queries on data about a play, a corpus or several corpora
    \item \textbf{terminological variation}: queries can be in (more or less) “normal” language or use specialised terms from drama research or literary theory
    \item \textbf{DraCor specificity variation}: DraCor itself has its own language: corpora have names (and abbreviations), plays have short titles (we call them “slugs”), and identifiers, values and features have their own labels etc.;  this DraCor-specific language can be used explicitly in queries
    \item \textbf{canonicity variation}: we expect an LLM to ‘know’ more about canonical texts than about little-known texts; we therefore vary our queries selectively so that we specifically ask queries about almost unknown texts\footnote{As a proxy for the classification as “non-canonical”, we take the existence or non-existence of an article on the play in the German-language Wikipedia. There are no Wikipedia articles on either “Der Nollhart” (experiment \hyperref[exp:1-5]{1-5}) or “Die entführte Dose” (experiment \hyperref[exp:5-4]{5-4} as of July 2025.}
\end{itemize}

\begin{table*}[!h]
\centering
\small
\renewcommand{\arraystretch}{1.5}
\caption{Experiments}
\label{tab:experiments}
\begin{tabular}{|p{0.8cm}|p{6.5cm}|p{1.2cm}|p{1.9cm}|p{1.9cm}|p{1.9cm}|}
\hline
\textbf{ID} (Set-No.) & \textbf{Prompt} & \textbf{Scope} & \textbf{Terminology Variation} & \textbf{DraCor Specificity Variation} & \textbf{Canonicity Variation} \\
\hline
1-1 & What is the number of characters in Dantons Tod? & play & 0 & 0 & 0 \\
\hline
1-2 & What is the number of dramatis personae in Dantons Tod? & play & 1 & 0 & 0 \\
\hline
1-3 & What is the number of characters in Dantons Tod in GerDraCor? & play & 0 & 1 & 0 \\
\hline
1-4 & What is the number of characters in Dantons Tod (buechner-dantons-tod) in GerDraCor? & play & 0 & 1 & 0 \\
\hline
1-5 & What is the number of characters in Der Nollhart? & play & 0 & 0 & 1 \\
\hline
2-1 & What is the mean number of characters in French Drama? & corpus & 0 & 0 & - \\
\hline
3-1 & Which drama corpus has the highest mean number of characters? & corpora & 0 & 0 & - \\
\hline
3-2 & Which drama corpus covers the widest time range? & corpora & 0 & 0 & - \\
\hline
4-1 & How does the percentage of female speakers in German drama change over time? & corpus & 0 & 0 & - \\
\hline
4-2 & How does the mean percentage of female speakers in Swedish drama change over time? & corpus & 0 & 0 & - \\
\hline
4-3 & How does the gender distribution in Swedish drama change over time? & corpus & 1 & 0 & - \\
\hline
4-4 & How does the percentage of female speakers in ItaDraCor change over time? & corpus & 0 & 1 & - \\
\hline
5-1 & Who is the most important character in Emilia Galotti? & play & 0 & 0 & 0 \\
\hline
5-2 & Who is the protagonist in Emilia Galotti? & play & 1 & 0 & 0 \\
\hline
5-3 & Which character is quantitatively most dominant in Emilia Galotti? & play & 1 & 0 & 0 \\
\hline
5-4 & Who is the protagonist in Die entführte Dose? & play & 1 & 0 & 1 \\
\hline
\end{tabular}
\end{table*}

\noindent
Table \ref{tab:experiments} lists the queries that we have prompted during our experiments and notes which of the variations listed above are active in each instance. We have grouped the queries into 5 sets:

\begin{itemize}
    \item \textbf{set 1}: queries that target data for one play (5 variants)
    \item \textbf{set 2}: queries that target statistics for a corpus (1 variant)
    \item \textbf{set 3}: queries that target a comparison of statistics for multiple corpora (2 variants)
    \item \textbf{set 4}: queries that target the combination of data for one corpus (4 variants)
    \item \textbf{set 5}: queries that target the application of a literary concept (“protagonist”) on one play (4 variants)
\end{itemize}

For each experiment, first the query was prompted. The result of this first query was then evaluated. In addition, a documentation prompt was entered,\footnote{The text of the documentation prompt is: “Document the entire chat in a Markdown file. Include a header with relevant metadata (e.g., settings used). For each step, provide information about the tools and DraCor API endpoints involved. Do not summarize. Focus on documenting the steps taken and the data used to answer the initial question. Simply record the conversation and all tool usage and analyses performed.” – As our experiments show, there is still some room for optimisation here as part of the prompt engineering. In particular, the requested metadata should be specified more precisely and the format should be more formalised. There is currently still too much variance in the documentation generated.} which resulted in an artefact in the form of a Markdown file documenting the experiment.\footnote{Cf. the repository on\\ \url{https://github.com/dracor-org/dracor-mcp-evaluation/tree/0d64119d69563627e0af7b873bd5506c1784690f/2025\_cda-preprint\_documentation}} Each query was then run a further 4 times using the ‘Retry’ function (‘With no changes’). The outputs were analysed for the reliability score (see below).

A research discussion on the evaluation of MCP servers is just emerging. Recently, \cite{gao_2025_mcpradar} proposed a multi-dimensional benchmark with the MCP-RADAR, but the paper (and the associated code) - like almost all publications on MCPs – is currently still without review. For this reason, we orient our evaluation on the conceptual thoughts on LLM agent evaluation by \cite{vongthongsri_2025_llm}, as these were developed in close connection with the already published LLM benchmark framework DeepEval (provided by Confident AI).

For evaluating the tool usage of so called “tool calling agents”, Vongthongsri suggests looking at two aspects: “\textbf{Tool Correctness}, which determines whether the right tools were called, and \textbf{Tool-Calling Efficiency}, which evaluates whether the tools were used in the most efficient way to achieve the desired results” \citep{vongthongsri_2025_llm}. While these aspects focus specifically on tool use, with regard to the entire agentic workflow, the aspect of \textbf{Task Completion}, among others, also needs to be evaluated. Taking these aspects as a starting point, we define the following evaluation criteria: 

\begin{itemize}
    \item \textbf{Correct Answer}: If the LLM comes to an answer and this answer is correct, we confirm “task completion”. As a reference for evaluating the correctness of the answers, we have carried out the necessary queries and calculations ourselves.\footnote{Cf. the Jupyter Notebook by Henny Sluyter-Gäthje:\\ \url{https://github.com/dracor-org/dracor-mcp/blob/main/evaluation/MCP-Evaluation_Paper-Version.ipynb}}
    \item \textbf{Tool Correctness}: If the LLM decides to use a tool (tool decision) and also uses the tools with which the query can be handled, we attest “tool correctness”.
    \item \textbf{Tool-Calling Efficiency}: For some queries there are several ways of processing. It also happens that the LLM first chooses a path that does not lead to the result and then looks for a new path. We evaluate this behaviour as “tool-calling efficiency”. If the LLM chooses the shortest processing path, we assign the value ‘5 = perfect’. We deduct one point for each dead end and for each more complicated processing path up to the value ‘1 = very bad’.
\end{itemize}

In view of the volatility in the choice of processing paths noted during testing and the subsequent unreliability of the tool use, we add reliability score.

\begin{itemize}
    \item \textbf{Tool-Use Reliability}: As outlined above, we repeated the first experiment, which formed the basis of the evaluation, four times (retry with no changes). If the basic solution (Tool Correctness) and the result (Correct Answer) are identical, we certify a reliable repetition. The maximum value is therefore 5/5 (always the same general processing path and the same answer after 5 attempts)
\end{itemize}

In the following, we document each experiment individually. At the end, we provide a summarised overview in Table \ref{tab:results}.

\subsection{Number of Characters of a Play}

\begin{experiment}{1-1 What is the number of characters in Dantons Tod?}
\label{exp:1-1}
\expitem{Comment on the Query}{scope: play; one data point; can be processed by using two tools}
\expitem{Tool Chain Used}{\code{DraCor:get\_corpora} – \\ \code{DraCor:get\_plays\_in\_corpus\_by\_title\_helper} – \\ \code{DraCor:get\_play\_characters}}
\expitem{Correct Answer}{yes (“103”)}
\expitem{Tool Correctness}{yes}
\expitem{Tool-Calling Efficiency}{4 (minus 1 for not using the information provided by\\ \code{DraCor:get\_plays\_in\_corpus\_by\_title\_helper}\\ but using \code{DraCor:get\_play\_characters} as an additional tool)}
\expitem{Tool-Use Reliability}{5/5}
\expitem{Documentation}{\\ \url{https://github.com/dracor-org/dracor-mcp-evaluation/blob/0d64119d69563627e0af7b873bd5506c1784690f/2025\_cda-preprint\_documentation/1-1\_DraCor-MCP-Paper\_2025-07.md}}
\expitem{Notes}{The model uses internal "knowledge" to pick the German drama corpus. The model does not understand the value for "network size", which can be taken from the response received by \code{DraCor:get\_plays\_in\_corpus\_by\_title\_helper}, as a value for the number of characters. So it counts the list containing the play's characters in the response of the \code{/corpora/{corpusname}/plays/{playname}/characters} endpoint provided by the tool.}
\end{experiment}

\begin{experiment}{1-2 What is the number of dramatis personae in Dantons Tod?}
\label{exp:1-2}
\expitem{Comment on the Query}{ scope: play; one data point; using terminology variation (“dramatis personae”); can be processed by using two tools}
\expitem{Tool Chain Used}{\code{DraCor:get\_corpora} – \\ \code{DraCor:get\_plays\_in\_corpus\_by\_title\_helper} – \\ \code{DraCor:get\_play\_characters}}
\expitem{Correct Answer}{yes (“103”)}
\expitem{Tool Correctness}{yes}
\expitem{Tool-Calling Efficiency}{4 (minus 1 for not using the information provided by\\ \code{DraCor:get\_plays\_in\_corpus\_by\_title\_helper}\\ but using \code{DraCor:get\_play\_characters} as an additional tool)}
\expitem{Tool-Use Reliability}{5/5}
\expitem{Documentation}{\\ \url{https://github.com/dracor-org/dracor-mcp-evaluation/blob/0d64119d69563627e0af7b873bd5506c1784690f/2025_cda-preprint_documentation/1-2_DraCor-MCP-Paper_2025-07.md}}
\expitem{Notes}{Same as \hyperref[exp:1-1]{1-1}}
\end{experiment}

\begin{experiment}{1-3 What is the number of characters in Dantons Tod in GerDraCor?}
\label{exp:1-3}
\expitem{Comment on the Query}{scope: play; one data point; using DraCor specificity variation (“GerDraCor”); can be processed by using one tool}
\expitem{Tool Chain Used}{\\ \code{DraCor:get\_plays\_in\_corpus\_by\_title\_helper} – \\ \code{DraCor:get\_play\_characters}}
\expitem{Correct Answer}{yes (“103”)}
\expitem{Tool Correctness}{yes}
\expitem{Tool-Calling Efficiency}{4 (minus 1 for not using the information provided by\\ \code{DraCor:get\_plays\_in\_corpus\_by\_title\_helper}\\ but using \code{DraCor:get\_play\_characters} as an additional tool}
\expitem{Tool-Use Reliability}{5/5}
\expitem{Documentation}{\\ \url{https://github.com/dracor-org/dracor-mcp-evaluation/blob/0d64119d69563627e0af7b873bd5506c1784690f/2025\_cda-preprint_documentation/1-3\_DraCor-MCP-Paper\_2025-07.md}}
\expitem{Notes}{Same as \hyperref[exp:1-1]{1-1}}
\end{experiment}

\begin{experiment}{1-4 What is the number of characters in Dantons Tod (buechner-dantons-tod) in GerDraCor?}
\label{exp:1-4}
\expitem{Comment on the Query}{scope: play; one data point; using DraCor specificity variation (“GerDraCor”; “buechner-dantons-tod”); can be processed by using one tool}
\expitem{Tool Chain Used}{\code{DraCor:get\_play\_characters}}
\expitem{Correct Answer}{yes (“103”)}
\expitem{Tool Correctness}{yes}
\expitem{Tool-Calling Efficiency}{5}
\expitem{Tool-Use Reliability}{1/5}
\expitem{Documentation}{\\ \url{https://github.com/dracor-org/dracor-mcp-evaluation/blob/0d64119d69563627e0af7b873bd5506c1784690f/2025\_cda-preprint\_documentation/1-4_DraCor-MCP-Paper\_2025-07.md}}
\expitem{Notes}{Strangely enough, Claude makes counting errors when repeating the processing of the query: Although it obtains the same character list with 103 entries as in the first call, it returns the results “101”, “93”, “100”, “93” in the four repetitions. According to the LLM, the cause of the problem is that it did not actually count, but answered the question using its internal ‘knowledge’ despite having the data from DraCor.\footnote{The specific wording of the response is: “The cause of my error was that I made an assumption and stated a number without actually counting the entries in the character list. I should have either 
	1. Counted the entries systematically, or 
	2. Used code to count them programmatically. 
Thank you for the correction - this is a good reminder to be more careful with data analysis and to verify counts rather than making assumptions. The DraCor API returned a complete list of all 103 characters in the play, and I failed to count them accurately.”}}
\end{experiment}

\newpage

\begin{experiment}{1-5 What is the number of characters in Der Nollhart?}
\label{exp:1-5}
\expitem{Comment on the Query}{scope: play; one data point; canonicity variation; can be processed by using two tools}
\expitem{Tool Chain Used}{\code{DraCor:get\_corpora} – \\ \code{DraCor:get\_plays\_in\_corpus\_by\_title\_helper} – \\ \code{DraCor:get\_play\_metadata}}
\expitem{Correct Answer}{yes (“14”)}
\expitem{Tool Correctness}{yes}
\expitem{Tool-Calling Efficiency}{4 (minus 1 for not using the information provided by\\ \code{DraCor:get\_plays\_in\_corpus\_by\_title\_helper}\\ but using \code{DraCor:get\_play\_metadata} as an additional tool)}
\expitem{Tool-Use Reliability}{5/5}
\expitem{Documentation}{\\ \url{https://github.com/dracor-org/dracor-mcp-evaluation/blob/0d64119d69563627e0af7b873bd5506c1784690f/2025\_cda-preprint\_documentation/1-5\_DraCor-MCP-Paper\_2025-07.md}}
\expitem{Notes}{Claude uses a similar approach to \hyperref[exp:1-1]{1-1}, but relies on a different tool (\code{DraCor:get\_play\_metadata}) to retrieve data on the characters. Interestingly, a second tool is used, because the result of the documentation prompt suggests that Claude already considered the value of the network size as an indicator of the number of characters (“Network size: 14 (indicating 14 characters)”), but still relied on counting the objects in the response of the second tool used to arrive at its final correct answer.}
\end{experiment}

\subsection{Mean Number of Characters of a Corpus}

\begin{experiment}{2-1 What is the mean number of characters in French Drama?}
\label{exp:2-1}
\expitem{Comment on the Query}{scope: corpus; two data points; can be processed by using one tool}
\expitem{Tool Chain Used}{\code{DraCor:get\_corpus} – \\ \code{DraCor:get\_minimal\_data\_of\_plays\_of\_corpus\_helper} – \\ \code{DraCor:get\_corpus\_metadata} – \\ \code{DraCor:get\_corpus\_metadata\_paged\_helper} – \\ \code{DraCor:get\_corpora}}
\expitem{Correct Answer}{yes (“9.19”)}
\expitem{Tool Correctness}{yes}
\expitem{Tool-Calling Efficiency}{3 (minus 1 for the choice of a more complicated processing path; -1 for the fact that the more complicated path leads to a dead end)}
\expitem{Tool-Use Reliability}{2/5}
\expitem{Documentation}{\\ \url{https://github.com/dracor-org/dracor-mcp-evaluation/blob/da9d4230bdb36584469bb26eec40469e2694a27e/2025\_cda-preprint\_documentation/2-1-1\_DraCor-MCP-Paper\_2025-07.md}}
\expitem{Notes}{FreDraCor is currently the largest corpus in DraCor, which poses some problems. The easiest way to process the query is to use \code{DraCor:get\_corpora} to load basic corpus metadata, including the number of plays and the number of characters in the entire corpus. In the end, this is also the path that the model chooses. Before that, it tries to fetch the corpus metadata via various metadata tools, but either fails due to the file size (“Data exceeded maximum length limit (1,048,576 characters)”) or cancels the plan to batch process the paged metadata. It is astonishing, though, that in the experiment documented, the processing path using the information in \code{DraCor:get\_corpora} is used after all. However, the choice of this path is not very stable: this path was only chosen in 2 out of 5 cases. More frequently, Claude chooses a sampling approach as an alternative processing path. In an experiment also documented here, using this way (and the corresponding tool chain) Claude came to the following conclusion based on a sample of 320 plays: “The mean number of characters in French drama is approximately 7.44 characters per play.”\footnote{\url{https://github.com/dracor-org/dracor-mcp-evaluation/blob/da9d4230bdb36584469bb26eec40469e2694a27e/2025_cda-preprint\_documentation/2-1-2\_DraCor-MCP-Paper\_2025-07.md}}}
\end{experiment}

\subsection{Comparison of Mean Number of Characters and Time Range of Publication Dates in Different Corpora}

\begin{experiment}{3-1 Which drama corpus has the highest mean number of characters?}
\label{exp:3-1}
\expitem{Comment on the Query}{scope: corpora; 26 (corpora) x 2 data points; can be processed by using one tool}
\expitem{Tool Chain Used}{\code{DraCor:get\_corpora}}
\expitem{Correct Answer}{yes (“GerShDraCor: 39.39 characters per play”)}
\expitem{Tool Correctness}{yes}
\expitem{Tool-Calling Efficiency}{5}
\expitem{Tool-Use Reliability}{5/5}
\expitem{Documentation}{\\ \url{https://github.com/dracor-org/dracor-mcp-evaluation/blob/da9d4230bdb36584469bb26eec40469e2694a27e/2025\_cda-preprint\_documentation/3-1\_DraCor-MCP-Paper\_2025-07.md}}
\expitem{Notes}{Nothing to note. Straightforward and reliable way of processing the query.}
\end{experiment}

\begin{experiment}{3-2 Which drama corpus covers the widest time range?}
\label{exp:3-2}
\expitem{Comment on the Query}{scope: corpora; 26 (corpora) x 2 data points; can be processed with 27 API calls}
\expitem{Tool Chain Used}{\code{DraCor:get\_corpora} – \code{DraCor:get\_corpus\_metadata} – \\ \code{DraCor:get\_corpus\_metadata\_paged\_helper} –\\ \code{DraCor:get\_minimal\_data\_of\_plays\_of\_corpus\_helper} –\\
\code{DraCor:get\_minimal\_data\_of\_plays\_of\_corpus\_helper} –\\
\code{DraCor:get\_minimal\_data\_of\_plays\_of\_corpus\_helper} –\\
\code{DraCor:get\_minimal\_data\_of\_plays\_of\_corpus\_helper} –\\
\code{DraCor:get\_minimal\_data\_of\_plays\_of\_corpus\_helper} –\\
\code{DraCor:get\_minimal\_data\_of\_plays\_of\_corpus\_helper} –\\
\code{DraCor:get\_minimal\_data\_of\_plays\_of\_corpus\_helper} –\\
\code{DraCor:get\_minimal\_data\_of\_plays\_of\_corpus\_helper} –\\
\code{DraCor:get\_minimal\_data\_of\_plays\_of\_corpus\_helper} –\\
\code{DraCor:get\_minimal\_data\_of\_plays\_of\_corpus\_helper} –\\
\code{DraCor:get\_minimal\_data\_of\_plays\_of\_corpus\_helper}}
\expitem{Correct Answer}{partly, i.e. 0.5 (answer is “French corpus spans 727 years”; “French” is correct, “727” is wrong – “847” is the correct answer)}
\expitem{Tool Correctness}{partly, i.e 0.5 (it uses the right tool, but uses it improperly)}
\expitem{Tool-Calling Efficiency}{3 (minus 2 for two dead ends, one due to response size errors, the another one due incorrect tool use)}
\expitem{Tool-Use Reliability}{2/5}
\expitem{Documentation}{\\ \url{https://github.com/dracor-org/dracor-mcp-evaluation/blob/da9d4230bdb36584469bb26eec40469e2694a27e/2025\_cda-preprint\_documentation/3-2\_DraCor-MCP-Paper\_2025-07.md}}
\expitem{Notes}{While this query can be processed in a simple Python script with 13 lines of code, for example, it is practically unanswerable for Claude in the current setup. There are two reasons for this: Claude cannot process the metadata files (\code{DraCor:get\_corpus\_metadata}) for medium and large corpora due to their size (“Response exceeded maximum length (1,048,576 characters”); at the same time, Claude refuses to use batch processing to process all pages of the paged metadata that it could retrieve for medium and large corpora using \code{DraCor:get\_minimal\_data\_of\_plays\_of\_corpus\_helper}. After these two approaches proved to be dead ends for Claude itself, it again chooses a sampling approach, whereby it quickly switches to always retrieving the first page and the last page using \code{DraCor:get\_minimal\_data\_of\_plays\_of\_corpus\_helper}. In doing so, it makes the assumption that the data sets are ordered chronologically, i.e. that the oldest play can be found on the first page and the newest play on the last page. However, this is a false assumption; the paginated list is not in chronological order. As a result, the sample created by Claude does not capture the time range of the corpus, or at best only by chance. In addition, it does not query the metadata for all corpora. Accordingly, the result is wrong. On top of this, Claude regularly chooses other ways of processing that also do not lead to correct results. So the reliability of the chosen approach is also low.}
\end{experiment}

\subsection{Percentage of Female Speakers over Time}

\begin{experiment}{4-1 How does the percentage of female speakers in German drama change over time?}
\label{exp:4-1}
\expitem{Comment on the Query}{scope: corpus; 737 (plays) x 2 data points; can be processed using two tools}
\expitem{Tool Chain Used}{\code{DraCor:get\_corpus} – \\ \code{DraCor:get\_corpus\_metadata} – \\ \code{DraCor:get\_minimal\_data\_of\_plays\_of\_corpus\_helper} – \\ \code{DraCor:get\_play\_characters} – \code{DraCor:get\_play\_characters} – \\ \code{DraCor:get\_play\_characters} – \code{DraCor:get\_play\_characters} – \\ \code{DraCor:get\_play\_characters}}
\expitem{Correct Answer}{no (see documentation)}
\expitem{Tool Correctness}{yes}
\expitem{Tool-Calling Efficiency}{ 4 (minus 1 for the metadata dead end)}
\expitem{Tool-Use Reliability}{2/5}
\expitem{Documentation}{\\ \url{https://github.com/dracor-org/dracor-mcp-evaluation/blob/da9d4230bdb36584469bb26eec40469e2694a27e/2025\_cda-preprint\_documentation/4-1\_DraCor-MCP-Paper\_2025-07.md}}
\expitem{Notes}{Claude's problem of reading the metadata files for medium and large corpora via \code{DraCor:get\_corpus\_metadata} is known from the previous experiments (“Response exceeded maximum length”). We have also already noted that Claude then unfortunately regularly chooses a more or less informed sampling approach that is never valid from a scholarly point of view. In this case, Claude uses \code{DraCor:get\_play\_characters} to retrieve data for five selected plays. The result is inadequate considering the variety of plays in the corpus.}
\end{experiment}

\begin{experiment}{4-2 How does the mean percentage of female speakers in Swedish drama change over time?}
\label{exp:4-2}
\expitem{Comment on the Query}{scope: corpus; 68 (plays) x 2 data points; can be processed using two tools}
\expitem{Tool Chain Used}{\code{DraCor:get\_corpus} - \code{DraCor:get\_corpus\_metadata}}
\expitem{Correct Answer}{yes (see documentation)}
\expitem{Tool Correctness}{yes}
\expitem{Tool-Calling Efficiency}{5}
\expitem{Tool-Use Reliability}{5/5}
\expitem{Documentation}{\\ \url{https://github.com/dracor-org/dracor-mcp-evaluation/blob/da9d4230bdb36584469bb26eec40469e2694a27e/2025\_cda-preprint\_documentation/4-2\_DraCor-MCP-Paper\_2025-07.md}}
\expitem{Notes}{Nothing to note. Straightforward and reliable way of processing the query.}
\end{experiment}

\begin{experiment}{4-3 How does the gender distribution in Swedish drama change over time?}
\label{exp:4-3}
\expitem{Comment on the Query}{scope: corpus; 68 (plays) x 2 data points; terminology variation (“gender”); can be processed using two tools}
\expitem{Tool Chain Used}{\code{DraCor:get\_corpus} – \code{DraCor:get\_corpus\_metadata}}
\expitem{Correct Answer}{yes (see documentation)}
\expitem{Tool Correctness}{yes}
\expitem{Tool-Calling Efficiency}{5}
\expitem{Tool-Use Reliability}{5/5}
\expitem{Documentation}{\\ \url{https://github.com/dracor-org/dracor-mcp-evaluation/blob/da9d4230bdb36584469bb26eec40469e2694a27e/2025\_cda-preprint\_documentation/4-3\_DraCor-MCP-Paper\_2025-07.md}}
\expitem{Notes}{The question is more open than \hyperref[exp:4-2]{4-2} as it is not only tailored to female characters but to the comparison of the gender. It also does not specify the basis on which the gender distribution should be analyzed. Claude analyses the characters as well as the authors of the plays in SweDraCor. It also combines the two aspects by analysing how many female and male characters there are per female and male author. All in all, a straightforward and reliable way of processing the query.}
\end{experiment}

\begin{experiment}{4-4 How does the percentage of female speakers in ItaDraCor change over time?}
\label{exp:4-4}
\expitem{Comment on the Query}{scope: corpus; 157 (plays) x 2 data points; using DraCor specificity variation (“ItaDraCor”); can be processed using two tools}
\expitem{Tool Chain Used}{\code{DraCor:get\_corpora} – \code{DraCor:get\_corpus\_metadata} – \\ \code{DraCor:get\_minimal\_data\_of\_plays\_of\_corpus\_helper} – \\ \code{DraCor:get\_play\_characters} – \code{DraCor:get\_play\_characters} – \\ \code{DraCor:get\_play\_characters}}
\expitem{Correct Answer}{no (see documentation)}
\expitem{Tool Correctness}{yes}
\expitem{Tool-Calling Efficiency}{4 (minus 1 for not using the metadata file)}
\expitem{Tool-Use Reliability}{2/5}
\expitem{Documentation}{\\ \url{https://github.com/dracor-org/dracor-mcp-evaluation/blob/da9d4230bdb36584469bb26eec40469e2694a27e/2025\_cda-preprint\_documentation/4-4\_DraCor-MCP-Paper\_2025-07.md}}
\expitem{Notes}{Again, Claude refuses to process the metadata file. The workaround via \\ \code{DraCor:get\_minimal\_data\_of\_plays\_of\_corpus\_helper} and \\ \code{DraCor:get\_play\_characters} then makes sense, but it’s quite cumbersome and again Claude chooses a sampling approach that does not lead to reliable results. In the case of repetitions, Claude sometimes (i.e. three times) chooses a more efficient processing path in which it actually analyses the metadata file, but even then, not all plays in the corpus are taken into account. }
\end{experiment}

\subsection{Catch the Protagonist}

\begin{experiment}{5-1 Who is the most important character in Emilia Galotti?}
\label{exp:5-1}
\expitem{Comment on the Query}{scope: play; number of data points depends on the operationalisation; four tools could be relevant}
\expitem{Tool Chain Used}{\code{DraCor:get\_play\_characters} – \\ \code{DraCor:get\_play\_metrics}}
\expitem{Correct Answer}{yes (“Marinelli” – however, this depends on the interpretation, which Claude also states itself from time to time)}
\expitem{Tool Correctness}{yes}
\expitem{Tool-Calling Efficiency}{5}
\expitem{Tool-Use Reliability}{5/5}
\expitem{Documentation}{\\ \url{https://github.com/dracor-org/dracor-mcp-evaluation/blob/da9d4230bdb36584469bb26eec40469e2694a27e/2025_cda-preprint\_documentation/5-1\_DraCor-MCP-Paper\_2025-07.md}}
\expitem{Notes}{The model pulls data from endpoints that are important for the answer. It would be conceivable to obtain additional data via \code{DraCor:get\_play\_metadata} and \code{DraCor:get\_spoken\_text\_by\_characters}. This is not done. When it comes to the repetitions, in some cases Claude uses the \code{Dracor:get\_play\_network} tool either in addition to or as an alternative to \code{DraCor:get\_play\_metrics}.The result remains the same.}
\end{experiment}

\begin{experiment}{5-2 Who is the protagonist in Emilia Galotti?}
\label{exp:5-2}
\expitem{Comment on the Query}{scope: play; number of data points depends on the operationalisation; using terminology variation (“protagonist”); four tools could be relevant}
\expitem{Tool Chain Used}{\code{DraCor:get\_play\_characters}}
\expitem{Correct Answer}{yes (“Emilia” – however, this depends on the interpretation)}
\expitem{Tool Correctness}{yes}
\expitem{Tool-Calling Efficiency}{5}
\expitem{Tool-Use Reliability}{5/5}
\expitem{Documentation}{\\ \url{https://github.com/dracor-org/dracor-mcp-evaluation/blob/da9d4230bdb36584469bb26eec40469e2694a27e/2025_cda-preprint_documentation/5-2_DraCor-MCP-Paper_2025-07.md}}
\expitem{Notes}{In this case, Claude limits itself to the data provided by \\ \code{DraCor:get\_play\_characters}. Although this approach makes sense, it lacks dimensions that could be introduced by using other tools. Nevertheless, the answer is based on plausible arguments. The effect of terminological variation should be highlighted here. Claude reliably distinguishes between the most important character (which is primarily defined on a quantitative basis) and the protagonist, whose status is primarily based on qualitative arguments. Again, there is almost no variation in the repetitions.}
\end{experiment}

\begin{experiment}{5-3 Which character is quantitatively most dominant in Emilia Galotti?}
\label{exp:5-3}
\expitem{Comment on the Query}{scope: play; number of data points depends on the operationalisation; using terminology variation (“quantitatively most dominant”, cf. Pfister 1991: 165); four tools could be relevant}
\expitem{Tool Chain Used}{\code{DraCor:get\_play\_metadata} – \code{DraCor:get\_play\_metrics} – \code{DraCor:get\_spoken\_text\_by\_characters}}
\expitem{Correct Answer}{yes (“Marinelli”)}
\expitem{Tool Correctness}{yes}
\expitem{Tool-Calling Efficiency}{5}
\expitem{Tool-Use Reliability}{5/5}
\expitem{Documentation}{\\ \url{https://github.com/dracor-org/dracor-mcp-evaluation/blob/da9d4230bdb36584469bb26eec40469e2694a27e/2025_cda-preprint_documentation/5-3__DraCor-MCP-Paper_2025-07.md}}
\expitem{Notes}{In this case, Claude makes use of diverse data and thus endeavours to make a complex, multi-perspective judgement. The additional use of the tool \code{DraCor:get\_spoken\_text\_by\_characters} should be emphasised in particular, as it adds a further dimension that supports the argumentation. However, in the case of three repetitions, Claude restricts itself to the \code{DraCor:get\_play\_metrics} tool; in the case of one repetition, it also uses \code{Dracor:get\_play\_characters}. However, the answer remains identical, even if it is less extensively backed up with data.}
\end{experiment}

\begin{experiment}{5-4 Who is the protagonist in Die entführte Dose?}
\label{exp:5-4}
\expitem{Comment on the Query}{scope: play; number of data points depends on the operationalisation; using terminology variation (“protagonist”); canonicity variation; four tools could be relevant}
\expitem{Tool Chain Used}{\code{DraCor:get\_corpora} – \\ \code{DraCor:get\_plays\_in\_corpus\_by\_title\_helper} – \\  \code{DraCor:get\_play\_characters} – \code{DraCor:get\_play\_metadata}}
\expitem{Correct Answer}{yes (“Foppendorf”)}
\expitem{Tool Correctness}{yes}
\expitem{Tool-Calling Efficiency}{5}
\expitem{Tool-Use Reliability}{5/5}
\expitem{Documentation}{\\ \url{https://github.com/dracor-org/dracor-mcp-evaluation/blob/da9d4230bdb36584469bb26eec40469e2694a27e/2025_cda-preprint_documentation/5-4_DraCor-MCP-Paper_2025-07.md}}
\expitem{Notes}{A multi-perspective approach to determine protagonism; not all relevant tools are used. The result is nevertheless unquestionable.}
\end{experiment}

\section{Overall Results and Discussion}

Table \ref{tab:results} aggregates the results of the experiments.

In the end, the overarching interpretation of the results of the experiments depends on what you expect from a system like the one tested. One may be impressed that LLMs can now conduct autonomous research actions which include retrieval of necessary data through an MCP Server (or comparable architectures). But one may also expect such interactions, which always suggest to impart knowledge, to bring a maximum of correctness and reliability – which they do not. In fact, this seems to describe the current state of our system quite well: It seems to be a promise of an infrastructure of the future, but one that is currently still too limited, too unsound, too unwieldy to be considered an infrastructure. 

\begin{table*}[!h]
\centering
\small
\renewcommand{\arraystretch}{1.5}
\caption{Results}
\label{tab:results}
\begin{tabular}{|p{0.8cm}|p{6.5cm}|p{1.7cm}|p{1.7cm}|p{1.7cm}|p{1.7cm}|}
\hline
\textbf{ID} & \textbf{Prompt} & \textbf{Correct Answer} & \textbf{Tool\newline Correctness} & \textbf{Tool-Calling Efficiency} \newline 1 = very bad 5 = perfect & \textbf{Tool-Use Reliability} \newline (X out of 5)\\
\hline
\hyperref[exp:1-1]{1-1} & What is the number of characters in Dantons Tod? & 1 & 1 & 4 & 5/5 \\ 
\hline
\hyperref[exp:1-2]{1-2} & What is the number of dramatis personae in Dantons Tod? & 1 & 1 & 4 & 5/5 \\
\hline
\hyperref[exp:1-3]{1-3} & What is the number of characters in Dantons Tod in GerDraCor? & 1 & 1 & 4 & 5/5 \\
\hline
\hyperref[exp:1-4]{1-4} & What is the number of characters in Dantons Tod (buechner-dantons-tod) in GerDraCor? & 1 & 1 & 5 & 1/5 \\
\hline
\hyperref[exp:1-5]{1-5} & What is the number of characters in Der Nollhart? & 1 & 1 & 4 & 5/5 \\
\hline
\hyperref[exp:2-1]{2-1} & What is the mean number of characters in French Drama? & 1 & 1 & 3 & 2/5 \\
\hline
\hyperref[exp:3-1]{3-1} & Which drama corpus has the highest mean number of characters? & 1 & 1 & 5 & 5/5 \\
\hline
\hyperref[exp:3-2]{3-2} & Which drama corpus covers the widest time range? & 0.5 & 0.5 & 3 & 2/5 \\
\hline
\hyperref[exp:4-2]{4-2} & How does the mean percentage of female speakers in Swedish drama change over time? & 1 & 1 & 5 & 5/5 \\
\hline
\hyperref[exp:4-3]{4-3} & How does the gender distribution in Swedish drama change over time? & 1 & 1 & 5 & 5/5 \\
\hline
\hyperref[exp:4-4]{4-4} & How does the percentage of female speakers in ItaDraCor change over time? & 0 & 1 & 4 & 2/5 \\
\hline
\hyperref[exp:5-1]{5-1} & Who is the most important character in Emilia Galotti? & 1 & 1 & 5 & 5/5 \\ 
\hline
\hyperref[exp:5-2]{5-2} &
Who is the protagonist in Emilia Galotti? & 1 & 1 & 5 & 5/5 \\
\hline
\hyperref[exp:5-3]{5-3} & Which character is quantitatively most dominant in Emilia Galotti? & 1 & 1 & 5 & 5/5 \\
\hline
\hyperref[exp:5-4]{5-4} &
Who is the protagonist in Die entführte Dose? & 1 & 1 & 5 & 5/5 \\
\hline
\end{tabular}
\end{table*}

Let's look at the numbers: 13 out of 16 answers were correct. Incorrect answers were due to problems with processing large amounts of data (and, possibly, and erroneous implementation of a tool)\footnote{Cf. \url{https://github.com/dracor-org/dracor-mcp/issues/8}}, not due to incorrect processing paths (resp. incorrect tool usage).

11 out of 16 answers could be reliably repeated. This is good. However, it also indicates a considerable degree of volatility in certain query settings. And we have not yet been able to identify patterns as to when LLM tool use becomes unreliable - and when it is reliable. This is certainly a major problem when it comes to building future infrastructures for Computational Literary Studies, for example.

Tool correctness, in contrast, is high: in 15 out of 16 experiments, the LLM (also) used the correct tool. This does not mean that it hasn’t used the wrong tools on its processing path. At some point, however, it (also) used the correct tool. Even then, errors could still occur (see the scores for “Correct Answer”), since even if the LLM used the correct tool, this doesn't indicate that it used the tool correctly.

Only in half of the cases (in 8 out of 16 experiments) were we able to assign the maximum score of 5/5 for tool efficiency. The mean value is 4.375. The lowest value is 3. That is not so bad. In other words, it is not the case that the LLM is wildly trying out all available tools. However, it is not uncommon for it to initially use the wrong tool or choose a more complicated processing path. From an end user's point of view, it is particularly disappointing that the LLM does not seem to build up any specialised "tool knowledge". There is something annoying about watching the LLM repeatedly use a tool that is unsuitable for the task at hand - even though it has already done this incorrectly dozens, if not hundreds of times before. As long as dedicated “tool use knowledge” – i.e. the development of a memory with experiential knowledge of which tool is good or not for which task – does not soon become a component in LLM architectures, we believe that this problem can be addressed in particular through comprehensive and carefully curated docstrings (as descriptions of the possible uses of tools, their limitations, their skilful handling, their idiosyncrasies, and the tricks needed to use them wisely, etc.). We will come back to this idea of docstring engineering later.

While the lack of ‘tool use knowledge’ that we frequently noticed in our experimental setup could be an architectural problem of the current LLM systems, two other problems have their roots in the restrictions on performance of the actual LLM interfaces. These interfaces we currently use to interact with the major commercial LLMs are geared towards the average customer. As a result, the reason for most of the errors and problems in our experiments was, 

\begin{itemize}
    \item first, the very fast reaching of the "maximum length limit" for data and,
    \item second, the refusal of the LLM to process batch queries.
\end{itemize}

Both tasks are actually computationally trivial. The resources required are minimal. Basically, it would be trivial even for Claude to produce a code artifact that solves these tasks computationally. But LLMs, like Claude, are a commercial mass product that works with extremely severe limitations, especially in terms of computing capacity. This considerably restricts the LLM's ability to act as a research assistant. The tactics apparently (in whatever way) implemented in Claude to use a sampling approach for computationally intensive activities are inadequate, and this not only because Claude does not master any valid sampling methods out of the box, as far as we see.

In addition to these conclusions (some of which relate very generally to the current status of the LLMs), we can derive some further insights from the specific setup of our experiments. Here, the domain-specific tasks that our MCP server has to process needs to be mentioned. In order to better understand the way in which the MCP server processes these domain-specific tasks, we have made some variations to our queries. On these variations, we can make the following observations:

\begin{itemize}
    \item The use of terminology (e.g. “protagonist” or “gender”) makes a difference in individual cases, but in any case leads to different answers. The extent to which LLMs have a kind of terminological knowledge, also with regard to literary studies, is still to be researched (cf. \citealt{jannidis_2025_llm}).
    \item The use of DraCor-specific vocabulary increases the efficiency of tool use. Simply put, the LLM knows much faster what exactly is meant when using the "dialect of the resource". Here we see great potential for docstring optimisation. Furthermore, this observation indicates that working with MCP servers could also be conceptualized as a problem of translation involving specialised languages.
    \item Finally, it should be noted that when using the DraCor MCP server, there is no noticeable difference between processing queries for canonical plays and non-canonical plays. In contrast, it can be stated without the MCP server that at least in one experiment, queries on non-canonical plays yielded no results at all,\footnote{Please compare the following chat with the documentation of experiment \hyperref[exp:5-4]{5-4}:\\ \textbf{Query:} "Who is the protagonist in Die entführte Dose?"\\ \textbf{Answer:} "I'm not familiar with a work titled "Die entführte Dose" (The Kidnapped/Abducted Can/Tin). This doesn't appear to be a well-known German literary work or film that I can identify from my knowledge.”} whereas the LLM was able to provide substantial information on the play when the MCP server was switched on.
\end{itemize}

While these observations do indeed provide initial insights into the use of MCP servers with CLS resources and infrastructures, it remains fundamentally questionable how stable these findings will be in view of the extremely rapid development in the field of LLM architectures. Irrespective of this, the observations and findings also point to a whole series of desiderata. From an experimental point of view, this includes 1) the professionalisation of the experimental setup (use of Claude's API, use of a web-based MCP server, more transparent repetition of queries); 2) the further development of the documentation of the experiments, whereby a stricter formalisation should be implemented here, which, if possible, also enables the automated evaluation of the outputs; 3) the algorithmisation of the evaluation mentioned above, which can hardly be evaluated manually, especially where it becomes a more complex process analysis (“agentic chains”), and collective systematic benchmarking as practiced by NLP/AI community. Finally, 4) it would be interesting to run test strategies against the DraCor API, for example based on XML or TXT files of the plays and linked to the question of how text data behaves in relation to API responses.

\section{Outlook: Towards the Art of Docstring Engineering}

If we return from the experiments to what was there at the beginning, namely a rapid prototype of an MCP server for Computational Literary Studies resources developed on the basis of DraCor, then at the end of the experiments we must first and foremost realise that our prototype has the potential, but also the need, to be developed further, to be refined, to extended, to be improved. We will work on that in the next months.

From our point of view and based on the current state of performance of the systems we have tested, this will require something that we have – in the title of this paper – somewhat cheekily labelled as the “art of docstring engineering”. We have already indicated what we mean by this at the beginning of chapter \ref{sec:implementation}. We base this claim on a strong suspicion that docstrings, as stated in chapter \ref{sec:implementation}, “form a semantic contract between the tools and the LLM”. This suspicion arose in the course of building the MCP and conducting the experiments, namely that LLM almost exclusively uses the docstrings of the tools on our MCP server as the basis for deciding which tool(s) to use and the way in which to use them. Of course, the first task of the “docstring engineering” ahead of us is to empirically substantiate our suspicion. As a next step, we therefore envisage a series of experiments in which we use a “docstring variation method” to test what effect docstrings actually have on the tool usage of the LLM. 

If these tests confirm our suspicions, we currently expect that four design principles or challenges will be the focus of our “DraCor MCP Docstring Engineering”:

\begin{itemize}
    \item Docstring Engineering must both clearly identify the simple and unambiguous functions of MCP tools and keep options open for the creative (scholarly) use of these tools.
    \item It is also against this background that design approaches for docstrings need to be developed that characterise both the functional specificity of a tool as well as its possible embeddedness in agentic chains (‘workflows’?) in which it is connected to other tools, data, artefacts and scripts etc.
    \item Docstring Engineering has to systematically deal with the performance limitations of current LLM systems.
    \item It could make sense to document an expanding set of evaluated and reflected best practices (e.g. of entire workflows) for the use of a DraCor MCP-enabled LLM for our scholarly (and our non-scholarly) users – and via the MCP server also provide this documentation to the LLM as context, so that the LLM always already has ideas on how tools can be used meaningfully (and in the best case: creatively).
\end{itemize}

Docstring engineering might be the issue at hand that helps us tackle at least some of the current challenges in working with agentic systems.

In the end, we believe the potential is remarkable: agent-enabled LLMs, cleverly instructed by the art of docstring engineering, are already beginning to empower humanities scholars without coding skills to interact computationally with the resources of the literary tradition. It might be up to us to make sure that these agents stop talking nonsense.

\section*{Data Availability}

\begin{itemize}
    \item DraCor Data can be found here: \url{https://github.com/dracor-org}
    \item Documentation of the conversations with Claude can be found here: \url{https://github.com/dracor-org/dracor-mcp-evaluation/tree/main/2025\_cda-preprint\_documentation}
\end{itemize}

\section*{Software Availability}

\begin{itemize}
    \item The code repository for DraCor API can be found here: \url{https://github.com/dracor-org/dracor-api}
    \item DraCor MCP Server can be found here: \url{https://github.com/dracor-org/dracor-mcp} For our experiments in this paper, we used DraCor MCP Server \citep{boerner_2025_dracor-mcp}, v1.0.0-alpha.2, \url{https://github.com/dracor-org/dracor-mcp/releases/tag/v1.0.0-alpha.2}. \\
    Unfortunately, this version of our MCP server no longer works with the current versions of Claude Desktop (“Claude for Mac”, Version 0.12.55). For up-to-date compatibility information and installation instructions, consult the DraCor MCP Server repository README: \url{https://github.com/dracor-org/dracor-mcp}.
    \item Jupyter Notebook for validation can be found in this repo: \url{https://github.com/dracor-org/dracor-mcp-evaluation}

\end{itemize}

\section*{Acknowledgments}
The paper has benefited from “DraCorOS. Fostering Open Science in Digital Humanities”. We acknowledge the OSCARS project, which has received funding from the European Commission’s Horizon Europe Research and Innovation programme under grant agreement No. 101129751.
\\ \\ \noindent
The paper has also benefited from the project “Computational Literary Studies Infrastructure. CLS INFRA”, which has received funding from the European Union’s Horizon 2020 research and innovation programme under grant agreement No. 101004984.

\section*{Author Contributions}

    \textbf{Peer Trilcke}: Funding acquisition, Conceptualization, Methodology, Formal analysis, Investigation, Writing - Original Draft\\ \textbf{Ingo Börner}: Conceptualization, Software (DraCor MCP Server), Formal analysis, Investigation, Writing - Original Draft \\
    \textbf{Henny Sluyter-Gäthje}: Conceptualization, Formal analysis, Investigation, Writing - Original Draft \\
    \textbf{Daniil Skorinkin}: Conceptualization, Writing - Review \& Editing \\
    \textbf{Frank Fischer}: Funding acquisition, Writing - Review \& Editing \\
    \textbf{Carsten Milling}: Software (DraCor API)

%%
%% Define the bibliography file to be used
\bibliography{references}

\end{document}